\documentclass{appolb}
\usepackage{epsfig}

\newcommand{\ave}[1]{\left\langle #1 \right\rangle}

\newcommand{\eps}{\varepsilon}
\begin{document}
\title{Freeze-out by bulk viscosity driven instabilities
\thanks{Presented at Epiphany 2008, Krakow}}%
\author{Giorgio Torrieri
\address{Institut f\"ur Theoretische Physik,
  J.W. Goethe Universit\"at, Frankfurt am Main, Germany}
\and
Boris Tom\'a\v{s}ik
\address{Univerzita Mateja Bela, Bansk\'a Bystrica, Slovakia 
and Faculty of Nuclear Science and Physics Engineering, Czech Technical University, 
Prague, Czech Republic}
\and
Igor Mishustin
\address{Frankfurt Institute for Advanced Studies, Frankfurt am Main, Germany \\
and Kurchatov Institute, Russian Research Center, Moscow 123182, Russia.}}
\maketitle
\begin{abstract}

We describe a new scenario (first introduced in \cite{firstpaper}) for freezeout in  heavy ion collisions that could solve the lingering problems associated with the so-called HBT puzzle. We argue that bulk viscosity increases as $T$ approaches $T_c$.   The fluid than becomes unstable against small perturbations, and fragments into clusters of a size much smaller than the total size of the system.   These clusters maintain the pre-existing outward-going flow, as a spray of droplets, but develop no flow of their own, and hadronize by evaporation. We show that this scenario can explain HBT data and suggest how it can be experimentally tested.
\end{abstract}
  
\section{Introduction}
One of the unsolved issues in heavy ion physics, on both a fundamental and a phenomenological level, is the problem of particle emission from a strongly interacting system:  A wealth of experimental data seems to point to the conclusion that the ``matter'' created in heavy ion collisions is hot, continuus, and locally thermalized \cite{whitebrahms,whitephobos,whitestar,whitephenix}.
In the most energetic collisions, the local thermalization seems to happen so quickly as to give evidence that strongly interacting matter is a ``perfect fluid'', with a viscosity over entropy density ratio lower than for any other material.

In principle, it is not trivial to describe in detail how such a strongly coupled medium can dissolve and emit weakly coupled particles.  The most ``obvious'' picture is that of ``freeze-out'':  As the system cools down, the mean free path of the constituent particles increases, and, at a certain point, reaches a value comparable to the system size.   At this point, particles decouple.

The problem with this approach is that many details are unknown.  It is not well known how the hadronic mean free path changes in a hot dense medium.  It is not known at all what the degrees of freedom are, and how strongly interacting they are, in the vicinity of the phase transition.  These ambiguities have pushed the heavy ion community do adopt somewhat ad hoc formulae to describe the process of decoupling: The simplest approach consistent with energy conservation and ideal hydrodynamics is the Cooper-Frye formula \cite{cooperfrye}, assuming that at a certain critical spacetime surface (usually defined in terms of a critical temperature), the mean free path goes from zero to infinity.  An additional refinement is to use the Cooper-Frye distribution not as an output, but as an input into a hadronic kinetic model \cite{shuryak,hirano}.

It is not so surprising that these models fail to describe the interferometric particle measurements \cite{hbtpuzzle}, thought to indicate the spacetime distribution of the ``surface of last scattering'' \cite{hbtreview}.
The character of the data-model discrepancy is, however, interesting:

Measured parameters $R_{o}$ and $R_{s}$ (see section \ref{hbtclust} of this work or \cite{hbtreview} for a definition) are  nearly identical over all ranges of energy and system size.   
Their (positive) difference $R_{o}^2 - R_{s}^2$ is thought to 
correspond---somewhat simplified---to the duration of particle emission.
Hence, it looks like the fireball emits particles almost instantaneously and does not 
show any sign of phase transition or crossover. Hydrodynamics, with ``reasonable'' freeze-out condition (such as a critical temperature of 100 MeV or so) can not describe this even qualitatively.   This is more puzzling if one considers that a large difference between $R_{o}^2 - R_{s}^2$ was previously used as a signature \cite{predictions} of the onset of a phase transition, since the softening of the equation of state in the transition region would have greatly lenghtened the emission time.  Given the wide acceptance of the hypothesis that the degrees of freedom seen in heavy ion collisions are thermalized quarks and gluons, the lack of a firm explanation for interferometry data is puzzling.
Recently, RHIC HBT radii have been correctly described by a hydrodynamic model \cite{florkowski1,florkowski2}, where, however, the system is ``forced'' to freeze-out simultaneusly (particles stop interacting after formation) at a high ($>m_\pi$) temperature.  This calculation hints as that the missing physics might have to do with the reason hadrons seem to stop interacting at such a high temperature.  

Compounding this puzzle is the scaling of all HBT radii with
the multiplicity rapidity density $(dN/dy)^{1/3}$, over a large range 
of energies and system sizes \cite{lisa}. This scaling is typical for an isentropically expanding fluid that suddenly breaks apart.   While this is encouraging for practitioners of fluid dynamics applied to heavy ion collisions (albeit it suggests that the ``perfect fluid'' is not exclusive to RHIC energies), a dynamics that could break up the system instantaneusly and independently of energy is currently missing.

In this talk, we wish to suggest that this puzzle is linked to a well-known feature of QCD: Its approximate conformal invariance in the perturbative regime, combined with the presence of a non-perturbative conformal anomaly.  We argue that this suggests that the bulk viscosity of the system suddently spikes close to $T_c$, and that this could trigger instabilities that rapidly break the system into evaporating clusters.

\section{Bulk-viscosity driven clustering}

The QCD lagrangian with only light quarks is nearly conformally invariant: The only terms breaking conformal invariance are the light quark masses, which are small w.r.t. the other scales relevant to Quark-Gluon Plasma physics (temperature, energy density and so on).

It is therefore thought that dynamics of a QGP with no heavy quarks is also conformally invariant.   This means the pressure ($p$) is to a good approximation equal to a third of the energy density ($\epsilon$) and the bulk viscosity ($\zeta$) is much smaller than the entropy density ($s$) \cite{lifshitzlandau}.

Within the pQCD framework, this has been confirmed:
The bulk viscosity of high temperature strongly interacting matter has recently been calculated using perturbative QCD \cite{amybulk}, and found to be negligible, both in comparison to shear
viscosity and w.r.t. its effect on any reasonable collective evolution of the system.

In the hadron gas phase, of course, the numerous scales associated
with hadrons render conformal invariance a bad symmetry, and hence it
is natural to expect that bulk viscosity is not negligible. 
This is, again, rooted in a fundamental feature of QCD:  the
non-perturbative \textit{conformal anomaly}, that manifests itself in the
scale (usually called $\Lambda_{QCD}$) at which the QCD coupling
constant stops being small enough for the perturbative expansion to
make sense.  This scale approximately coincides with the scale at which confining forces hold hadrons together.

This violation of conformal invariance is not seen perturbatively, but
should dominate over the perturbatively calculated bulk viscosity as temperature drops
close enough to the QCD phase transition.

What happens to \textit{bulk} viscosity in this regime, where hadrons are not yet
formed, presumably the matter is still deconfined, but conformal
symmetry is badly broken?  While we can not as yet calculate this rigorously, 
there is compelling numerical evidence \cite{pratt,kharbulk,kkbulk,bulklattice} 
that bulk viscosity rises sharply, or even diverges, close to the phase transition temperature.

Because of the conformal simmetry argument above, this is not too surprising.
It becomes even less surprising if the character of the two phase transitions, deconfinement and chiral simmetry breaking, are examined more closely;
The the shear ($\eta$) and bulk ($\zeta$) viscosities 
roughly scale as \cite{hosoya,jeonvisc,weinberg}
\begin{eqnarray}
\label{bulkgeneral}
\eta & \sim & \tau_{\rm elastic} T^4 \\
\zeta & \sim & \left( \frac{1}{3} - v_s \right)^2 \tau_{\rm inelastic} T^4
\end{eqnarray}
where $\tau_{\rm (ine)elastic}$ refers to the equilibration timescale of (ine)elastic
collisions.

The dependence of $\tau_{\rm inelastic}$ on temperature can be guessed from
the fact that, at $T_c$, the quark condensate $\ave{q \overline{q}}$ 
acquires a finite non-zero value, and the gluon condensate increases.  Thus, a system exactly at $T_c$ will respond to any infinitesimal heating by creating $\ave{q \overline{q}}$ or gluon pairs,while any infinitesimal cooling will destroy them.
It is therefore clear that, if chiral simmetry were exact,  timescales of processes
that create extra $q \overline{q}$  pairs would diverge at $T_c$ analogously to correlation lenghts in other second order phase transitions.
As shown in \cite{jeonvisc}, bulk viscosity is sensitive to the timescale of such processes (the divergence has been directly checked in numerical simulations \cite{pratt}.

The sharp rise of bulk viscosity can also be understood within string kinetics:
confinement, microscopically, can be thought of
as a ``string tension'' appearing in the potential.
The appearance of such a string tension changes near-equilibrium kinetics profoundly even if the tension is very low compared to the typical momentum exchange, and therefore the relevant degrees of freedom are ``slightly confined'' quarks and gluons.
With no string tension, the vast majority of collisions in any reasonable kinetic model are elastic, and hence the diffusion across the trace of the energy momentum tensor (to which bulk viscosity is proportional to) is negligible.
With even a small string tension, {\em every} collision becomes inelastic.
Because of this, diffusion across the trace of the energy momentum tensor goes sharply up.

Could the sharp rise in bulk viscosity be the missing physics responsible for making hydrodynamic models agree with data?   For this to be the case, the bulk viscosity should trigger the system to decouple earlier, and faster, than the ``conventional'' prescription of freeze-out at a critical temperature would allow.

One mechanism which could lead to such a result are hydrodynamic instabilities.
As shown in  \cite{stability}, the stability condition of the Bjorken solution, provided a conformal equation of state holds, is\footnote{As noted in \cite{stability} the dependence of stability on the Reynolds number is {\em opposite} to that in usual non-relativistic hydrodynamics: Systems with small Reynolds numbers can be unstable.  A lower limit for the Reynolds number also exists, but it depends strongly on the wavelenght of the perturbations.  See \cite{stability} for more details}.
\begin{equation}
\eta + \frac{3}{4} \zeta < \frac{3}{4} s T \tau
\end{equation}
where $\tau$ is the proper time of the volume element, and $T$ is its temperature.
for the conformally invariant plasma \cite{cft} ($\eta = s/4 \pi,\zeta=0$) this requirement is automatically satisfied for a realistic start of the hydrodynamic evolution.

The sharp rise in bulk viscosity, however, makes the boost-invariant solution of the system linger for a long time in a constant temperature state.  This state, however, is unstable against small perturbations, which then have the time and possibility to blow up to a scale significantly altering the background evolution, and producing a highly inhomogeneus system. See \cite{stability_us} for a demonstration of this effect.

\begin{figure*}[tb]
\centerline{\epsfig{file=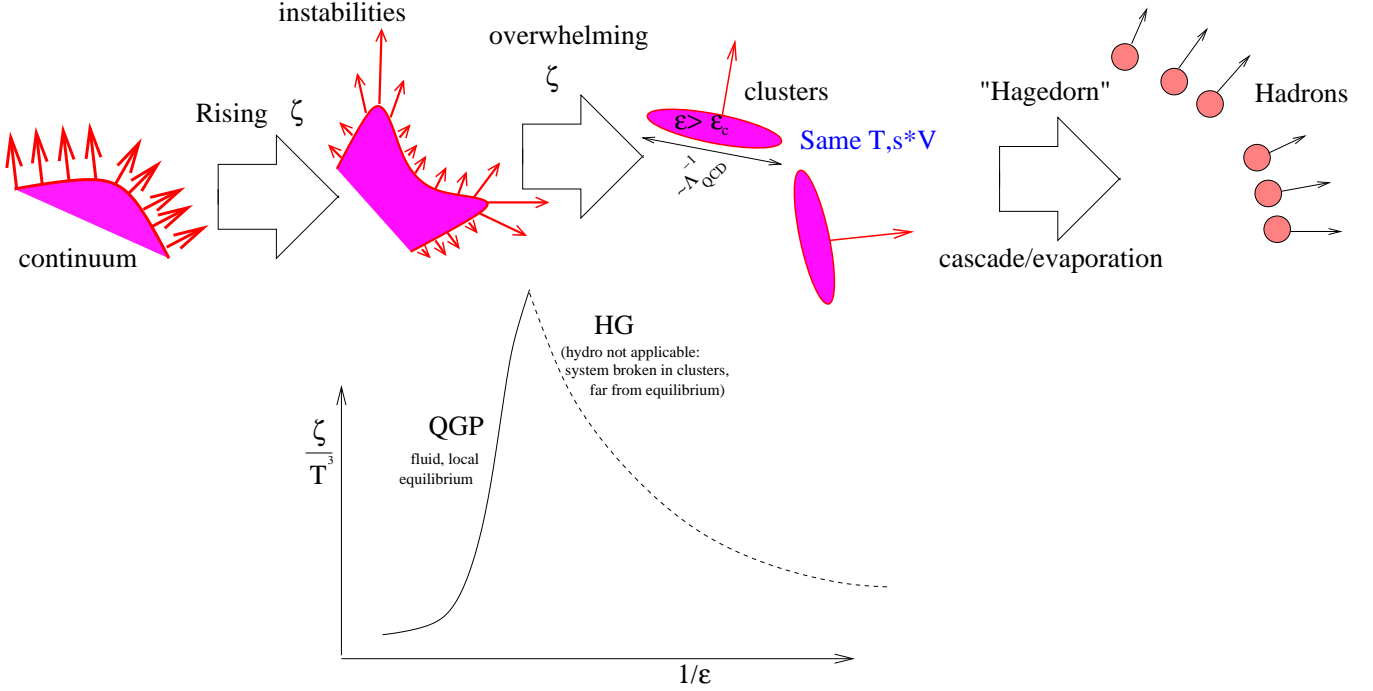,width=18cm}}
\caption{\label{diagram} (Color online) 
Fragmentation of the fireball due to sharply increasing bulk viscosity as
the temperature decreases. Matter which expanded easily before we describe as
oil. It suddenly becomes very rigid against expansion (described as honey in the 
figure) and breaks up into fragments. Hadrons evaporate from these fragments. 
}
\end{figure*}

The schematic evolution of the hot spots is then illustrated in Fig. \ref{diagram}: the effect of viscosity is to introduce a force resisting expansion and deformation.  This can not be accomplished globally, since causality prevents viscous forces from affecting the allready generated flow.  Viscous forces can however quench any expansion of the grown instability, rendering it ``rigid'' and disconnected from the rest of the system (the effective pressure being cancelled by the bulk viscosity).  

Each instability can then be considered as a stand-alone hot bubble, or cluster, moving with pre-existing flow.  It then presumably emits particles by evaporation.

Thus, we have reproduced a scenario similar to the bubble-nucleation picture
more commonly associated with supercooling \cite{kapusta,sudden1,sudden2,mish1,mish2,sudden3,randrup}.
This scenario, however, does not rely on the existance of a first-order phase transition.   Additionally, unlike \cite{kapusta}, the nucleation examined here does not result in an entropy increase, since  the formation of clusters should quickly kill off any gradient in the flow velocity ( $\partial_{\mu} u^\mu$), so entropy generation ($\sim (\partial_\mu u^\mu)^2$) during clustering should be negligible.
Thus, this model should obey the constraints, pointed out in \cite{etele} on multiplicity measurements.

The cluster size will be an interplay of local scales: $\Lambda_{QCD},T_c$ and the expansion rate $\partial_\mu u^{\mu}$.   For a rough estimate, we recall that
the energy momentum tensor, with vanishing shear viscosity but non-vanishing
bulk viscosity is
\begin{equation}
T^{\mu\nu} = (\eps + p) u^\mu u^\nu - p g^{\mu\nu} + \zeta\, \partial_\rho u^\rho \, (g^{\mu\nu} - u^\mu u^\nu)
\end{equation}
From energy-momentum conservation $\partial_\mu T^{\mu\nu} = 0$ we then obtain the rate of energy density
decrease
\begin{equation}
\label{erate}
\frac{1}{\eps}u^\mu \partial_\mu \eps = 
\frac{\eps+ p - \zeta \partial_\rho u^\rho}{\eps}  \partial_\mu u^\mu\, .
\end{equation}
Note that when $\zeta \partial_\rho u^\rho \sim p$ the energy density decreases at the same rate as 
if no work was performed in case with vanishing viscosity. For lower rates of the energy density decrease
the expansion even {\em decelerates}. Microscopically, this is mediated by inter-particle forces which
hold the system together. It can happen that the inertia of the bulk overcomes these forces and the 
system thus {\em fragments}. 

In order to obtain a more quantitative estimate of droplet size, we determine it
by the balance of deposited energy and collective expansion energy.
According to the definition of viscosity, it determines the amount of energy deposited per 
unit volume and unit time, i.e.
\begin{equation}
E_{\rm dis}=\int dV \int d\tau \zeta(\partial_{\mu} u^{\mu})^2,
\end{equation}
where $\zeta$ is bulk viscosity and $u^\mu$ collective 4-velocity. For simplicity let us assume again the Bjorken \cite{bjorken} picture. 
Then $\partial_\mu u^\mu=1/\tau$ and the 3-velocity is $v_z=z/t$. If bulk viscosity is indeed rapidly divergent at 
$T_c$, we can replace it with the $\delta$-function
\begin{equation}
\zeta(\tau)= 
\zeta_c T_c \delta \left( T(\tau) - T_c \right) = 
\zeta_c T_c \left. \frac{d \tau}{d T } \right|_{T=T_c} \delta(\tau-\tau'_c),
\end{equation}
where
$\zeta_c$ is a model parameter which should be given by deeper theoretical consideration.  If we call $\tau'_c =  T_c \left. \frac{d \tau}{d T } \right|_{T=T_c}$  we get
\begin{equation}
E_{\rm dis}=SL\frac{\zeta_c}{\tau'_c},
\end{equation}
where $S$ is the transverse area of the Bjorken cylinder and $L$ is the droplet longitudinal size. We consider a 
droplet whose center of mass is located at $z=0$ (though this assumption is not really important due to the 
boost invariance of the system).

The kinetic energy of droplet's expansion, which is in fact dissipated due to viscosity, is in non-relativistic limit
\begin{equation}
E_{\rm kin}=\frac{1}{2}\int dV\, \eps(\tau)v_z^2,
\end{equation}
where $\eps(\tau)$ is the  internal energy density of the fluid. It is of course a function of time but the above expression contains only volume integration. Let us evaluate the integral at the critical point, when actual break-up happens, then
\begin{equation}
E_{\rm kin}=\frac{S\, \eps_c}{24t_c^2}L^3.
\end{equation} Taking $t_c\approx\tau'_c$, we get finally
\begin{equation}
\label{dsize}
L^2=\frac{24\zeta_c\tau'_c}{\eps_c}\, .
\end{equation}
Notice that $\tau'_c$ in the numerator is actually the inverse expansion rate 
$\partial^\mu u_\mu$. Thus the droplet size squared is inversely proportional to 
the expansion rate. 
Within this scenario the droplet size will
grow with the lifetime of the hydrodynamic stage (from the initial temperature $T_0$ to $T_c$), but the growth will 
generally be 
slower than linear.
For our toy model example where the system has a conformal equation of state
 and boost-invariance ($dN/dy \sim \epsilon_0^{3/4} \sim \tau'_c$), this growth will be $\sim (dN/dy)^{1/2}$, but it 
is likely to be slower than that when  transverse expansion is considered.

Whether the cluster size is indeed only dependent on the internal scale of the system $\Lambda_{QCD}$ or on an interplay between the internal and collective scales (Eq. \ref{dsize}) is difficult to determine from first principles, as it depends on a quantitative understanding of the details of the non-equilibrium evolution around $T_c$.

The main point argued in the last section, one that does not depend on these
details, is that the sharp rise of bulk viscosity could force the system to break up into disconnected fragments, of a scale and lifetime much smaller than
the size of the system ($\mathcal{O}(1\phantom{A} \mathrm{ GeV}))$.  
These 
clusters 
then flow 
apart with 
pre-existing flow velocity and, presumably, decay by Hagedorn cascading.   In the next section we shall examine the effect this kind of freeze-out has on heavy ion phenomenology.



\section{Phenomenology of clustering}

There are two classes of observables where clustering can be naturally looked for:  momentum fluctuations/correlations, and particle interferometry.

\subsection{Clustering in event-by-event observables}
Forward-backward multiplicity correlations \cite{fbcorrel} and angular correlations in 
Cu+Cu collisions  at RHIC \cite{roland_darmstadt} have indicated the presence of clusters at freeze-out, the slow dependence of the clusters with the system size, and that the contribution of these clusters seems greater than that expected from just hadronic resonance decay.

 The scaling of $p_T$ fluctuations also provides direct evidence
that  particles are emitted from clusters, containing a small ($\sim 5$)
number of particles independently of collision energy or centrality \cite{ptfluct}.  The under-prediction, by the equilibrium statistical model, of fluctuations of ratios such as $K/\pi$ \cite{sqm2006} compounds this evidence, since cluster emission would enhance fluctuations of multiplicity yields and ratios.

A more direct signature of instabilities such as clusters is provided by the Kolomogorov-Smirnov test \cite{tomasks}:  If, at freeze-out, the system is entirely Cooper-Frye, than while each event will be different, the {\em probability density function} for observables will be the same (up to resonances and initial state fluctuations) for all momentum observables (rapidity, $p_T$,and azimuthal angle).   There is a statistically rigorous way to test the equivalence of two empirical distributions, the Kolmogorov-Smirnov test.  An analysis of heavy ion data using this method is in progress.  We hope that it will lead to a signature of event-by-event differences above those expected in hydrodynamics.

\subsection{Clustering in HBT \label{hbtclust}}
In HBT interferometry, the most usual coordinate system used is that defined in terms of the momentum of each particle pair and the beam:
In this usually used ``out-side-long'' coordinate system \cite{hbtreview},
$l$ (``long'') is the z direction (parallel to the
beam), $o$ (``out'') is the direction of the pair momentum, and $s$ (``side'') is the
cross product of the previous two.

In the Gaussian approximation (the correlation function of particle momenta is a Gaussian), HBT radii are directly related \cite{hbtreview} to
the system's correlation between the respective space directions $x_{o,s}$ and the emission time $t$
\begin{eqnarray}
R_s^2(K) &=& \ave{(\Delta x_s)^2} \label{rside}\\
R_o^2(K) &=& \ave{(\Delta x_o)^2} - 2 \frac{k_T}{k_0} \ave{\Delta x_o
  \Delta t}
    + \left(  \frac{k_T}{k_0} \right)^2  \ave{(\Delta t)^2} \label{rout}
\end{eqnarray}
where the $k$ vector is the sum of the two momenta (the first element, $k_0$, is $\simeq \sqrt{m^2 + \vec k^2}$).
For the most central events, because of cylindrical simmetry $x_o \sim x_s$.
The $R_o \sim R_s$ result is not easy to reconcile with naive hydrodynamics
plus a straight-forward (critical temperature) emission for two reasons:

First, the higher the initial energy, the larger the final system size, and the longer the emission time, and hence the expected discrepancy between $R_o$ and $R_s$.  
If the
system starts close to the mixed phase, the timescale of freezing out
should be longer still due to the softest point in the equation of state.
Hence, a generic prediction from Eqs. \ref{rside} and \ref{rout} is that 
$R_o/R_s>1$, broadly increases with energy, and exhibits a peak when the energy 
density is such that the system starts within, or slightly above the mixed phase.  This is in direct contrast with
experimental data, where $R_o/R_s \simeq 1$ is a feature at all reaction
energies.

In addition, generally, a fluid freezes out by both evaporation from a surface
and final decoupling as the system cools down.  In both cases the
$\ave{\Delta x \Delta t}$ correlation is negative, since particles on
the outer side are the first to freeze-out.   This increases
$R_o/R_s$ further (cf.\ eq~\ref{rout}).  
Time dilation due to transverse flow does not help
enough, as calculations show.   

Clusters can, in principle, help with both these issues. 
Cluster size, density and decay timescale, are approximately independent of either
reaction energy or centrality, as can be deduced from Eq.~\ref{dsize}. 
Hence, the near energy independence of the (comparatively short) emission timescale,
and hence of $R_{o}/R_{s}$, should be recovered.

If the decay products do not interact (or do not interact much)
after cluster decay, it can also be seen that $\ave{\Delta x \Delta t}$
can indeed be positive because of time dilation in cluster decay.

Recovering the  linear scaling of the radii with $(dN/dy)^{1/3} (\sim
N_{\rm clusters})$  \cite{lisa}, 
while maintaining the correct $R_{o}/R_{s}$ is also possible if
the clusters decay when their distance w.r.t.\ each other is
still comparable
to their intrinsic size.

Quantitative  calculations are necessary before determining whether these
constraints can be satisfied.  The technical details of how to perform such calculations, 
from a hydrodynamic code output with a critical temperature and cluster size, are outlined in the Appendix of \cite{firstpaper}.  
Hydrodynamics output is needed to specify the cluster flow array $u^\mu_i$ and emission array $\Sigma_\mu^i$, (the locus of spacetime points where the cluster formation occurs).   
The bulk-viscosity-driven freeze-out adds another parameter to ab initio HBT calculations:  
in addition to critical temperature/energy density, we now have the cluster size.
To see whether this helps solving the HBT problem, output from hydrodynamics with a high ($T \sim T_c$) freeze-out temperature should be fragmented into clusters with a certain distribution in size, 
which then produce hadrons according to the prescription in the Appendix of \cite{firstpaper}.

In conclusion, 
We have described a mechanism to generate fragments that is solidly grounded in QCD, and does not require a first order phase transition.
Hence, it is possible that hadronization is governed by this mechanism in all regimes where an approximately locally thermalized deconfined system is produced.

Future work in this direction including quantitative signatures of our model in both HBT data and event-by-event observables is in progress.


GT would like to thank the Alexander von Humboldt Foundation and Frankfurt
University for the support provided, and to Mike Lisa, Sangyong Jeon, Guy Moore and Johann Rafelski for fruitful discussions. 
BT acknowledges support from 
VEGA 1/4012/07 (Slovakia) as well as MSM 6840770039 and LC 07048
(Czech Republic).
IM acknowledges support provided by the DFG grant 436RUS 113/711/0-2
(Germany) and grants RFFR-05-02-04013 and NS-8756.2006.2 (Russia).
GT would like to thank the Institute For Nuclear Physics in Krakow for their hospitality and support during this conference, and W. Florkowski, W. Broniowski,M. Chojnacki and P. Bozek for discussions.

\end{document}